\documentclass[a4paper, 11pt]{article}

\usepackage{longtable}
\usepackage{geometry}
\usepackage{graphicx}
\usepackage{subfig}
\usepackage{color}
\usepackage{xcolor}
\usepackage{bm}
\usepackage{fixmath}
\usepackage{mathrsfs}
\usepackage{ulem}
\usepackage{multirow}
\usepackage{pdflscape}
\usepackage{rotating}
\usepackage{lscape}
\usepackage{pdflscape}
\usepackage{titlesec}
\usepackage{float}
\usepackage{hyperref}
\usepackage{eucal}
\usepackage{amsfonts}
\usepackage{amsmath}

\def\b{\begin{eqnarray}}
\def\e{\end{eqnarray}}

\begin{document}

\begin{center}
{\huge \textbf{Mathematical Analysis of the \vskip.3cm van der Waals Equation}}


\vspace {10mm}
\noindent
{\Large \bf Emil M. Prodanov}
\vskip.8cm
    {\it School of Mathematical Sciences, Technological University Dublin,} \\
    {\it Park House, Grangegorman, 191 North Circular Road, } \\
    {\it Dublin D07 EWV4, Ireland} \\
    {\it E-Mail: emil.prodanov@tudublin.ie}
\vskip1cm

\end{center}

\vskip2cm

\noindent
\begin{abstract}
\noindent
The parametric cubic van der Waals polynomial $p V^3 - (R T + b p) V^2 + a V - a b$ is analysed mathematically and some new generic features (theoretically, for any substance) are revealed --- if the pressure is not allowed to take negative values [temperatures not lower than $1/(4Rb)$], the localization intervals of the three volumes on the isobar--isotherm are: $3b/2 < V_A \le 3b$, $\,\, 2b < V_B < (3 + \sqrt{5})b$, and $3b \le V_C < RT/p + b = V_0 + b$ (with $V_0$ being Clapeyron's ideal gas volume). For lower values of the temperature, the root $V_A$ is bounded from below by $b$, while $V_B$ has the localization interval $b < V_B < 2a/(R \, \tau)$, where $\tau > 0$ is the new minimum temperature of the model. The unstable states of the van der Waals model have also been generically localized: they lie in an interval within the localization interval of $V_B$. A discussion on finding the volumes $V_{A, B, C}$, on the premise of Maxwell's hypothesis, is also presented.
\end{abstract}


\newpage


\noindent
\section{Introduction}
Johannes Diderik van der Waals was awarded Nobel Prize in 1910 ``for his work on the equation of state for gases and liquids". This work originated in his doctoral thesis from 1873 \cite{thesis}, where he put forward an ``equation of state" accommodating both the gaseous and liquid states of a substance and also demonstrated that these two states are of the same nature and merge continuously into each other: the van der Waals equation predicts critical behaviour and the experimentally observed transition between gas and liquid. There are two factors due to which a non-dilute aggregate of moving particles fails to comply with the ideal gas law: the {\it proper volume} of the particles (one has to account for the volume $b$ of the molecules, leading to less available volume, $V-b$, for their motion) and the so-called {\it internal pressure} $a/V^2$ (with $a > 0$), due to the mutual attraction of the gas molecules. Each of these two factors is associated with a correcting term, hence generalizing the ideal gas law  $p V = RT$ of Clapeyron to
\b
\label{vdw1}
\left( p + \frac{a}{V^2} \right) (V - b) = RT.
\e
Van der Waals never expected that in his equation, $a$ and $b$ would be constants \cite{nobel}, so that the equation would agree numerically with the experiments. As far back as 1873, he emphasized the possibility that $a$ might vary with the temperature (as $a \sim 1/T$) and also recognized the variability of $b$ with $V$ and not with $T$ \cite{nobel}. Later, he suggested that, even with constant $b$, the van der Waals equation is valid for a single substance and that many phenomena would be explained qualitatively provided that suitable values for $a$ and $b$ were introduced for mixtures \cite{nobel}. For the purposes of the mathematical analysis of the van der Waals equation in this paper and for the determination of some of its generic features (applying, theoretically, to all substances), the parameters $a$ and $b$ will be taken as two positive constants. \\
It is the richer mathematical structure of the {\it cubic equation} that allows, when used as equation of state, to accommodate critical behaviour and phase transitions,
and the van der Waals equation is the first proposed such equation (it is cubic in the volume $V$). It has been followed by over 200 different cubic equations of state aimed at studying various phase phenomena in multi-component mixtures (many of these equations of state are modifications of the original van der Waals equation) --- see \cite{anderko} for the study of equation-of-state methods for the modelling of phase equilibria, \cite{valderrama} for a review on cubic equations of state, and the references therein.  The van der Waals equation has been extensively studied and applied to various models for almost 150 years --- see the review \cite{kontogeorgis} and the significant reference list in it. However, as cubic equation, it is mathematically fairly complex and its mathematical aspects, including how to solve the equation, are rarely, if ever, discussed. \\
Without loss of generality, when in numerical form, the coefficients of the van der Waals equation can be considered to be integer numbers. This can be seen in the following way. Any measurement has finite precision and hence, irrational or transcendental numbers cannot be recorded in result of measurement. Measured quantities, as the pressure $p$ and the temperature $T$, which enter the van der Waals equation as parameters, are represented by numbers with finite number of digits after their decimal points. The parameters $a$ and $b$, as well as the universal gas constant $R$, are also such numbers (obtained in their own right via measurements with finite precision). Hence, the coefficients of the van der Waals equation, which are functions of all these, are also such numbers (within, of course, the chosen level of accuracy). Representing them as rational numbers and multiplying across by the highest denominator (some power of $10$), renders all coefficients of the van der Waals equation integer. \\
The  cubic in $V$ van der Waals equation, $p V^3 - (R T +b p) V^2 +  a V - ab = 0$, can be solved only point-wise in the parameter space, that is, for fixed values of $p$, $T$, $a$, and $b$ --- the roots can be found either numerically or by using the explicit Cardano formul\ae \, for the cubic (after depressing the equation --- eliminating the quadratic term --- by a suitable coordinate transformation). The van der Waals polynomial is an irreducible one in general, namely, it is not possible to express a solution of this cubic by radicals with radicands in the field of the real numbers: when an irreducible cubic polynomial is present, that is, when all of its three roots are real and distinct, and not one of them is rational (which, by the rational root theorem, would be in the form of an integer factor of the integer free term $ab$ divided by an integer factor of the integer  leading coefficient $p$), getting the roots by radicals involves taking cube roots of complex-conjugate numbers. The latter appear due to the necessity to involve imaginary numbers in the cubic Cardano formulae by taking the square root of a negative number. This is the so-called {\it casus irreducibilis} --- termed so in the 16th century, as it was not possible back then to take roots of negative numbers (the thinking, at the time, was “geometrical” – numbers had to be positive, as they represented lengths, areas, etc. – negative numbers were avoided). For temperatures below the critical temperature $T_{cr}$, the van der Waals equation  in general falls into the realm of {\it casus irreducubilis} --- by Maxwell's hypothesis (which is also referred to as Maxwell's construction), it has three distinct positive real roots, and the existence of a rational one among them is not guaranteed. \\
It is even more difficult to obtain a solution of the van der Waals equation when it is in symbolic form, i.e. without having the equation parameters fixed at certain values --- the {\it casus irreducibilis} forces the consideration of square roots of complex functions. There is also a complication arising due to Maxwell's hypothesis --- the one and same pressure $\widetilde{p}$, at which, during the isothermic expansion or contraction at any temperature $\widetilde{T} < T_{cr}$, the volume $V$ takes the value of any of the three real and distinct roots $V_{A, B, C}$ of the van der Waals equation, is immediately determined as soon as the temperature $\widetilde{T}$ of the isotherm on the $p\!\!-\!\!V$ diagram is fixed --- and to solve the van der Waals equation, one has to also uncover the relationship between $\widetilde{p}$ and $\widetilde{T}$. \\
The aim of this work is to address certain mathematical aspects of the {\it symbolic} van der Waals equation (when the parameters are not fixed) and to reveal features which are common for all substances, for all pressures and temperatures. As the roots of the {\it symbolic} equation cannot be determined, this paper finds the localization intervals of the roots of the van der Waals equation: following ideas developed in \cite{23}--\cite{32} for the mathematical analysis of polynomials of different degrees, it is found that, when the van der Waals equation has three real roots $V_A < V_B < V_C$, they satisfy $3b/2 < V_A \le 3b$, $\,\, 2b < V_B < (3 + \sqrt{5})b$, and $3b \le V_C < RT/p + b = V_0 + b$, where $V_0$ is Clapeyron’s ideal gas volume. This result is generic and it stems from the structure of the van der Waals equation and not from physical considerations (except imposing positivity on the pressure, namely, temperatures not lower than $1/(4Rb)$, and lower bound $b$ of the volume). The unstable states of the van der Waals model are also generically localized: they lie within the localization interval of $V_B$. \\
For temperatures below $1/(4Rb)$, the pressure $p$ on the $p\!-\!V$ diagram will exhibit a ``dip" below the abscissa. Then, root $V_A$ will be bounded from below by $b$, the root $V_B$ will have the localization interval: $b < V_B < 2a/(R \, \tau)$, where $\tau > 0$ is the new minimum temperature of the model. The unstable states of the van der Waals equation fall into this interval too.


\section{The van der Waals Equation and Maxwell's Hypothesis}
The van der Waals equation is often written as
\b
\label{vdw2}
p(V) \,\, = \,\, \frac{RT}{V-b} - \frac{a}{V^2} \,\, = \,\, \frac{RTV^2 - aV + ab}{V^2(V-b)}
\e
or as
\b
\label{vdw}
p V^3 - (R T +b p) V^2 +  a V - ab = 0.
\e
At first, the pressure $p$ will be strictly positive [it will be shown that for $p > 0$ one needs temperature $T > T_0 = a/(4Rb)$]. The volume $V$ will be strictly greater than $b$ (it cannot be smaller than the volume of the molecules). The graph of $p(V)$ for different $T$ is shown on Figure 1. \\
Equation (\ref{vdw}) is a cubic polynomial in the volume $V$. In addition to the model parameters $a$ and $b$ ($R$ is the universal gas constant), one can consider $p$ and $T$ as just two further parameters with their variation leading to different roots $V$ for this equation.  \\
Maxwell assumes that the second law of thermodynamics must be valid even when unstable states intervene between the initial and final states \cite{b}. For the van der Waals equation, the unstable states are those for which the pressure decreases with the decrease of the volume, namely, those between the local minimum (point $M$) and the local maximum (point $N$) of $p(V)$ --- see Figure (1). Applying $T dS = dU + p dV$ to the cycle, involving such unstable states, from point $A$, via points $M$, $B$, and $N$ to point $C$ along the isotherm (with the segment from $M$ via $B$ to $N$ consisting entirely of unstable points) and then returning from point $C$, via point $B$, to point $A$ along the isobar (see Figure 1), yields $T \! \oint \! dS = \oint \! U + \oint\!  p \, dV$. Given that, over any cycle, $\oint \! dS = 0 = \oint \! dU$, one gets $\oint \! p \, dV = 0$ and, hence, the two areas bounded by the isotherm and the isobar must be equal --- see \cite{b} and problem 48 in {\cite{baz}. This means that the isobar, on which points $A$, $B$, and $C$ lie, connects the two phases $A$ and $C$, which are in contact at equilibrium. If this is not satisfied, then the two phases $A$ and $C$ cannot be in equilibrium, even though they lie on the same isotherm and the same isobar \cite{b}. Maxwell's hypothesis therefore dictates that there should be a constraint between $\widetilde{p}$ --- the same pressure at which $V_{A,B,C}$ simultaneously exist --- and the temperature $\widetilde{T}$ of the isothermal contraction or expansion. That is, if the temperature $T$ is fixed at some $\widetilde{T}$, then $\widetilde{p}$ is immediately determined. Only then the van der Waals equation would yield three volumes that satisfy the Maxwell hypothesis.
\begin{figure}
\centering
\includegraphics[height=8.8cm, width=0.58\textwidth]{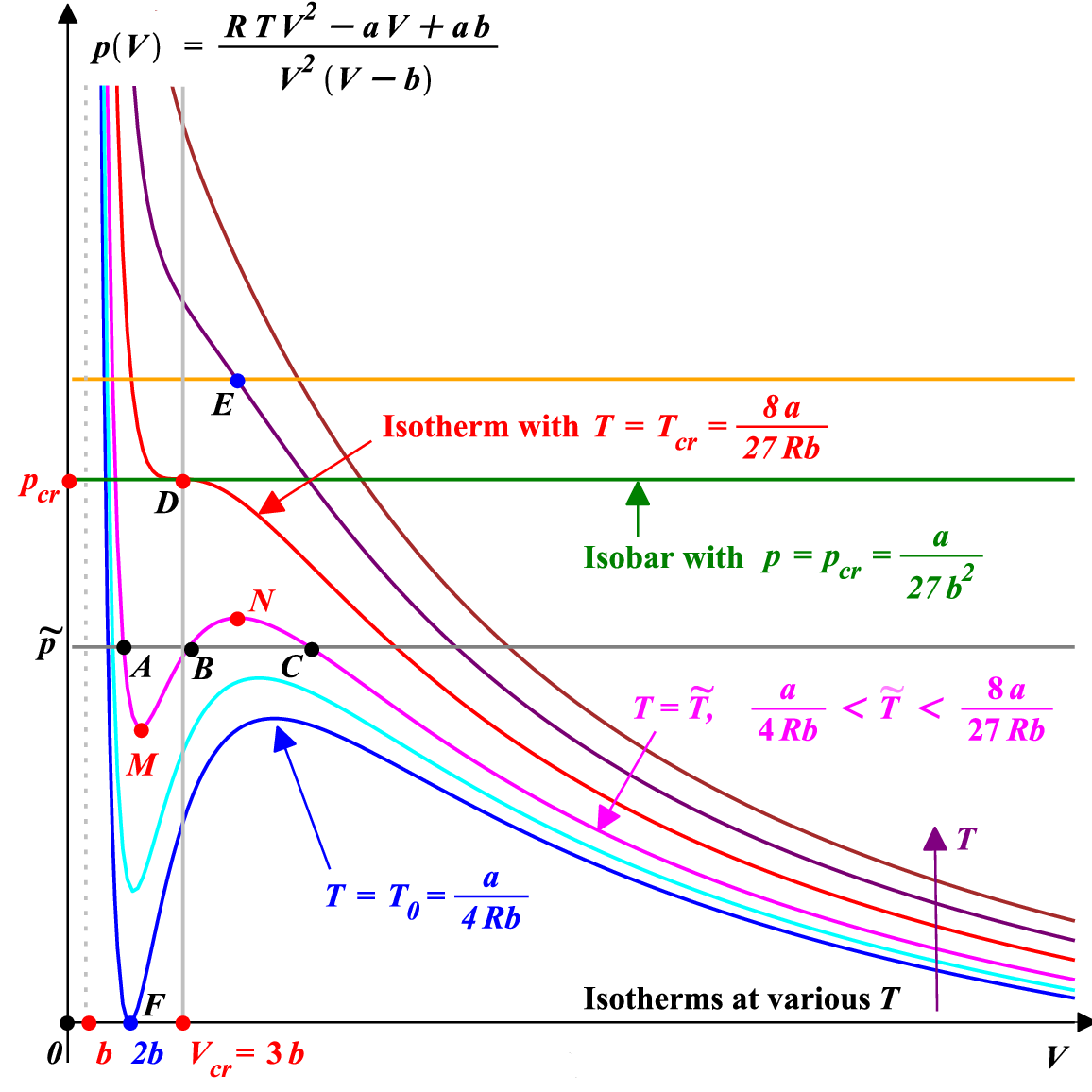}}
{\begin{minipage}{36em}
\scriptsize
\vskip.3cm
\begin{center}
{\bf Figure 1} \\
\vskip.3cm
{\bf The Van der Waals equation of state} $\bm{p(V) = \frac{R T V^2 - a V + a b}{V^2 (V - b)}}$  {\bf for various temperatures} \\
\end{center}
\end{minipage}
\end{figure}


\noindent
For two different isotherms at the same volume, the one at the higher temperature has higher pressure (there is one and only one isotherm through each point, that is, the isotherms do not intersect). Moving across the isotherms, which foliate the $p\!-\!V$ space, by increasing the temperature, points $A$ and $C$ on the isotherm--isobar get closer to each other in such manner that, at all times, the two areas bounded by the isobar and the isotherm remain equal. At temperature equal to the critical temperature $T_{cr}$, points $A$, $B$, and $C$ coalesce at point $D$. The volume at that point, $V_D$, is called critical volume $V_{cr}$. Thus, there is a triple root $V_{cr} = V_A = V_B = V_C$ when $T = T_{cr}$. The value of the pressure at this point is called critical pressure $p_{cr}$. The intersection of the isobar $p = p_{cr}$ with the isotherm $T = T_{cr}$ is at the inflection point $V_{cr}$ of the isotherm and the tangent of the isotherm at the inflection point is the isobar $p = p_{cr}$ itself (parallel to the abscissa). Macroscopically, the system represents a single phase at this state \cite{baz}. For $T < T_{cr}$, point $A$ corresponds to liquid and is stable, while point $C$ corresponds to gaseous state of the given substance and is meta-stable. [Namely, it is in internal equilibrium --- within the range of configurations accessible by continuous changes with the system having the lowest possible free energy. Under a large fluctuation --- the nucleation of a more stable phase --- transformation to the new phase would occur \cite{baz}.] Point $B$ is unstable and cannot be realized in nature. Upon the isothermal compression of a gas, once volume $V_C$ is reached, the evolution on the $p\!-\!\!V$ diagram no longer follows the isotherm. Some of the gas becomes liquefied and the two phases, gas (point $C$) and liquid (point $A$), exist simultaneously in equilibrium at the same temperature and pressure \cite{baz}. Further compression (beyond volume $V_A$) liquefies all of the gas. \\
A very essential part in the realization of this construct is the fact that the three distinct roots $V_{A, B, C}$, existing for $T < T_{cr}$, coalesce simultaneously into one triple root $V_{cr}$ at $T = T_{cr}$.

\section{Localization Intervals of the Roots of the van der Waals Equation}
If the discriminant $\Delta = -27 A^2 D^2 + 2(9 A C - 2 B^2)B D - 4 A C^3 + B^2 C^2$ of the general cubic polynomial $A x^3 + B x^2 + C x + D$ is negative, then there will be one real root and a pair of complex-conjugate roots. If $\Delta$ is positive, there will be three distinct real roots. If the discriminant is zero, the cubic will have a multiple root: a double root and a single root, when $B^2 > 3 A C$, or a triple root given by $-B/3A$, when $B^2 = 3 A C$  \cite{23}--\cite{32}. Sufficient condition for negative discriminant of the general cubic is $B^2 < 3 A C$  --- for such cubic, the Siebeck--Marden--Northshield equilateral triangle, which projects onto the three real roots, will not exist \cite{23}--\cite{32}. \\
All this can be seen as follows [see \cite{23}--\cite{32} for details]. The discriminant $\Delta$ is quadratic in the free term $D$ and the discriminant with respect to $D$ of the discriminant $\Delta$ is $\Delta_D = 16 (B^2 - 3AC)^3$. When $B^2 - 3AC < 0$, in view of the negative coefficient of the term with $D^2$ in $\Delta$, the cubic discriminant $\Delta$ will be negative and the cubic will have only one real root. When $B^2 - 3AC = 0$, the cubic discriminant $\Delta$, seen as a quadratic in $D$, will have a double root  $D = B^3/(27A^2)$, that is, the discriminant $\Delta$ will be zero for $B^2 - 3AC = 0$ and $D = B^3/(27A^2)$ [then $D$ will also be equal to $C^2/(3B)$]. In this case, the cubic will have a triple root $-B/3A$ (also equal to $-C/B$). However, if $D \ne B^3/(27A^2)$ while $B^2 - 3AC = 0$, the cubic discriminant $\Delta$ will not be zero --- it will be negative. Finally, when $B^2 - 3AC > 0$, the general cubic will have a positive discriminant between the roots of the quadratic in $D$ equation $\Delta = 0$, that is, between
$D_2 = [9ABC - 2B^3 - 2 (B^2 - 3AC)^{3/2}]/(27A^2)$ and $D_1 = [9ABC - 2B^3 + 2 (B^2 - 3AC)^{3/2}]/(27A^2)$. There will be three distinct real roots then. When $B^2 - 3AC > 0$  and $D = D_{1,2}$, the cubic discriminant $\Delta$  will  be zero, resulting in a double root and a single one. When  $B^2 - 3AC > 0$ and $D$ is outside of the roots $D_{1,2}$, the cubic discriminant $\Delta$  will be negative and there will be only one real root. \\
Under variation of the parameters of the equation, the discriminant can, eventually, change its sign. In the most general situation, as already seen, when the discriminant of the cubic is zero, one does not necessarily have $B^2 - 3 AC = 0$. This means the following. As, under variation of the parameters of the cubic, the discriminant approaches zero from below, there is one real root and a pair of complex-conjugate roots whose imaginary parts get smaller and smaller. The two complex-conjugate roots coalesce into a double real root when the discriminant becomes zero. When the discriminant becomes positive, this double real root bifurcates into two distinct real roots, the distance between which increases with the increase of the discriminant. For the van der Waals equation, there is a very distinctive feature: it has a triple root $V_{cr}$ at the critical temperature $T_{cr}$ and critical pressure $p_{cr}$. This is ``built into" the van der Waals equation by Maxwell's hypothesis: should there be a double root and a single root, rather than a triple root, Maxwell's hypothesis would fail as one of the two areas would then be zero. Hence, when the discriminant (which depends on $p$ and $T$, as well as $a$ and $b$) approaches zero, $B^2 - 3 AC$ also approaches zero. The van der Waals discriminant is
\b
\label{delta}
\Delta & \!\!\!\!\! = \!\!\!\!\! & - 4a \left[ b^4 p^3 + (3 R T b^3 + 2 a b^2)p^2 + (3 R^2 T^2 b^2 - 5 R T a b + a^2) p + R^2 T^2 \left( R T b - \frac{a}{4} \right)\!\right] \!. \nonumber \\
& & \phantom{emp}
\e
It is not simple to analyze as it is cubic in $p$ and cubic in $T$. \\
The term $B^2 - 3 AC$ for the van der Waals polynomial is
\b
\label{crit}
b^2 p^2 + (2 R T b - 3a) p + R^2 T^2.
\e
Both polynomials (\ref{delta}) and (\ref{crit}) are simultaneously zero if one takes $T = T_{cr} = (8 a)/(27 R b)$ and $p = p_{cr} = a/(27 b^2)$. These critical values of the temperature and the pressure are well known and will be confirmed further, following a different analysis. \\
Note that Maxwell's hypothesis prevents the van der Waals equation from having a double root (one of the two areas will be zero in such case). If a multiple root is present, this could only be the triple root $V_{cr} = 3b$ when $T = T_{cr}$ and $p = p_{cr}$. \\
Consider the van der Waals equation in the form (\ref{vdw2}). One can immediately determine the temperature range for having $p > 0$ in it. The graph of $p(V)$ exhibits a local minimum and a local maximum for temperature below the critical. With the decrease of the temperature, the ``dip" gets closer to the abscissa. Setting $p(V) = 0$, yields the quadratic equation $R T V^2 - a V + ab = 0$. The roots of this equation are real when the discriminant $-4abRT + a^2$ is positive, that is, when $T < T_0 = a/(4Rb)$. In such case, the curve $p(V)$ will cross the abscissa at $V_{1,2} = (a \pm \sqrt{-4abRT + a^2})/(2RT)$ and the pressure will be negative between $V_2$ and $V_1$. If the discriminant is zero [i.e. $T = T_0 = a/(4Rb)$], the curve $p(V)$ will have a double root at $V = a/(2RT_0) = 2b$. If the pressure is not allowed to become zero for a finite volume or become negative (at any volume), this discriminant must be negative and thus $T > T_0 = a/(4Rb)$ --- the minimum model temperature (otherwise the van der Waals equation would be ``pushed" too far). Boltzmann \cite{b} gives examples of isotherms entering a regime with negative pressure. One of them is mercury gradually extracted from a barometer tube. The other one is the isotherm at room temperature of distilled water --- it also dips below the abscissa. Discussion of such isotherms and the localization of the roots and extrema of the van der Waals equation in such case are presented further in this work.  \\
There is another characteristic temperature associated with the van der Waals equation. Note that, as the temperature increases, the curve on the $p\!-\!V$ diagram resembles the curve of an ideal gas with equation of state $p V = RT$. The question is, at what temperature the two could be identified. Equating $p V = RTV/(V - b) - a/V$ from the van der Waals equation to $RT$ from the Clapeyron law, one gets $T = a(V-b)/(VRb) = [a/(Rb)](1 - b/V)$. If the real gas is to resemble an ideal gas, then the molecular volume $b$ must be much smaller than $V$. Hence, one can take $b/V \to 0$ when seeking to emulate ideal gas behaviour. In this limit, one gets $T = T_b = a/(Rb) = 4T_0 = (27/8) T_{cr}$. This temperature is called Boyle's temperature. \\
To confirm that the critical temperature is indeed $8a/(27Rb)$ and that the critical pressure and volume are $a/(27b^2)$ and $3b$, respectively, consider the van der Waals polynomial $w(V) \equiv p V^3 - (RT + bp)V^2 + aV - ab$ from (\ref{vdw}) and re-write it as $V^3 w^\ast(V^{\ast})$ where
\b
\label{w*}
w^{\ast}(V^{\ast}) =  -ab V^{\ast^3} + a V^{\ast^2} - (R T + b p) V^\ast + p
\e
and $V^{\ast} = 1/V$ [this can be done as the free term $ab$ in the van der Waals polynomial $w(V)$ is not zero and thus 0 is not its root]. Hence, the roots of the van der Waals equation are the reciprocals of the roots of $w^\ast(V^\ast)$. The reason for the passage to the reciprocal polynomial $w^\ast(V^\ast)$ becomes apparent when one writes the equation $w^\ast(V^\ast) = 0$ as $- ab V^{\ast^3} + a V^{\ast^2} = (R T + b p) V^\ast - p$. Then the roots of $w^\ast(V^\ast)$ are the intersection points of $- ab V^{\ast^3} + a V^{\ast^2}$ and $(R T + b p) V^\ast - p$. With this trick, the parameters $T$ and $p$ were ``transferred" from the cubic to the straight line. The cubic $- ab V^{\ast^3} + a V^{\ast^2}$ is now ``fixed" by the parameters $a$ and $b$ only. By varying $p$, the $y$-intercept of the straight line $(R T + b p) V^\ast - p$ varies, as well as its slope. While variation of the temperature only affects its slope. This makes the analysis much simpler (as opposed to having to analyse a cubic with curvature depending on both $p$ and $T$, as would have been the case, had one split the original van der Waals equation in a similar manner).

\begin{center}
\begin{tabular}{cc}
\includegraphics[width=67mm]{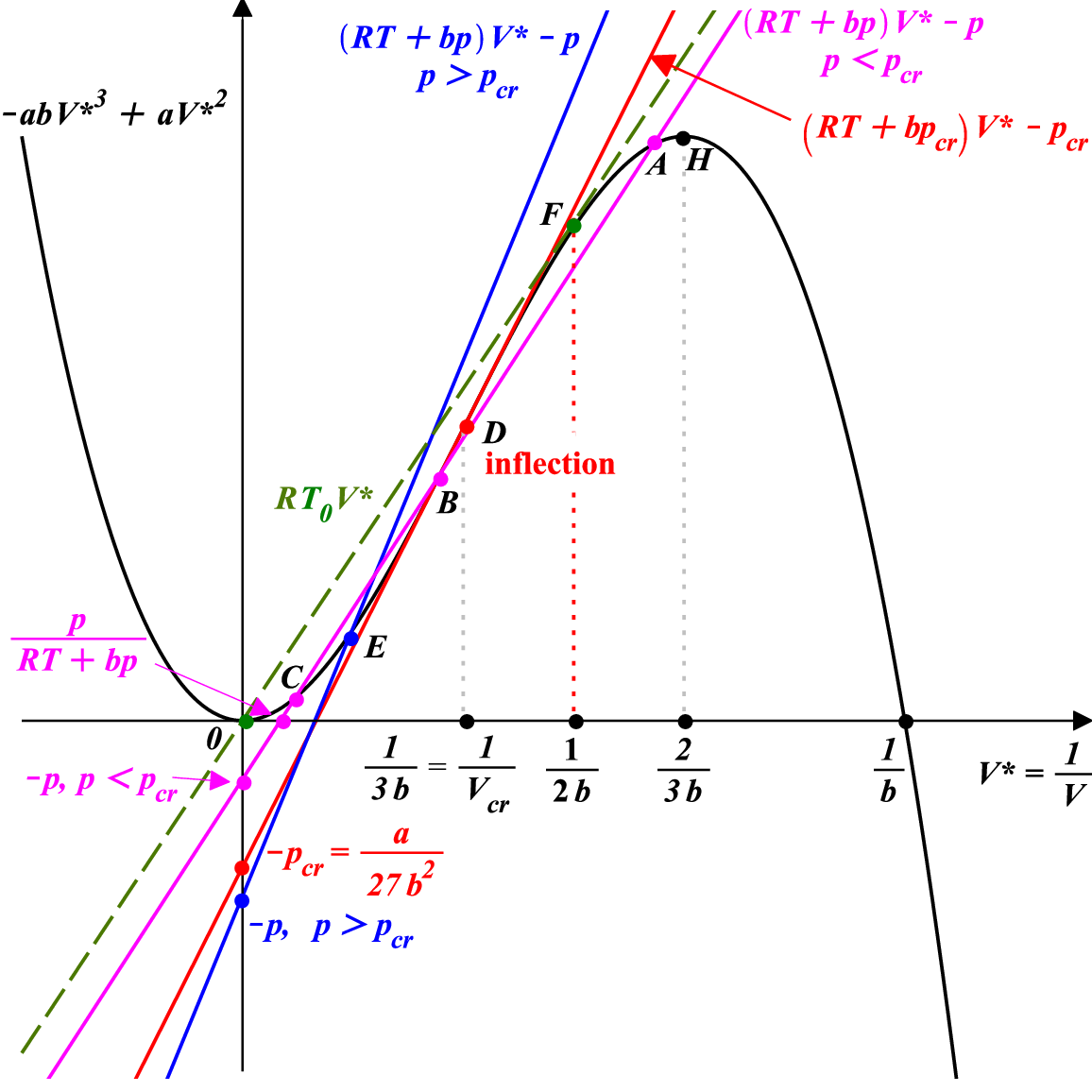} & \includegraphics[width=67mm]{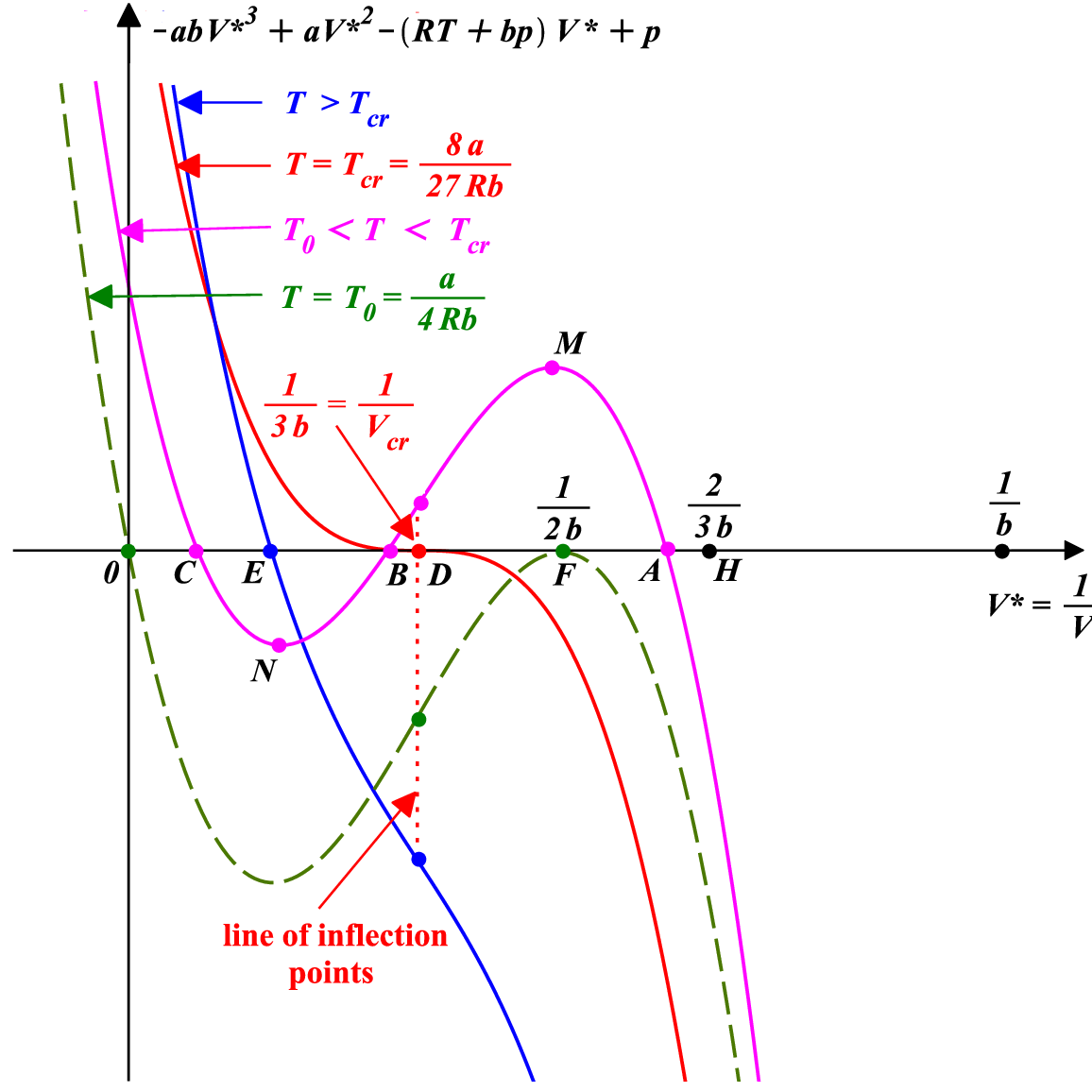} \\
{\scriptsize {\bf Figure 2a}} &  {\scriptsize {\bf Figure 2b}} \\
{\scriptsize {\bf The equation reciprocal to the Van der Waals}} & {\scriptsize {\bf The polynomial reciprocal to the van der Waals}} \\
{\scriptsize {{\bf equation:} \bm{$\!\!-ab V^{\ast^3} + a V^{\ast^2} = (RT + bp) V^\ast - p$}}} & {\scriptsize {{\bf polynomial:} \bm{$\!\!-ab V^{\ast^3} + a V^{\ast^2} - (RT + bp) V^\ast + p$}}} \\
& \\
\end{tabular}
\end{center}

\noindent
Introduce:
\b
w^{\ast}_L(V^{\ast}) & = &  -ab V^{\ast^3} + a V^{\ast^2}, \\
w^{\ast}_R(V^{\ast}) & = &  (R T + b p) V^\ast - p.
\e
Hence, $w^{\ast}(V^{\ast}) = w^{\ast}_L(V^{\ast}) - w^{\ast}_R(V^{\ast})$. The graphs of $w^{\ast}_{L,R}(V^{\ast})$ are shown on Figure 2a and that of $w^{\ast}(V^{\ast})$ --- on Figure 2b. The inflection point $w^{\ast}_L(V^{\ast})$ (point $D$ on the graphs) depends only on the equation parameters $a$ and $b$ and occurs at $V^\ast = 1/(3b)$. This coincides with the inflection of the whole $w^{\ast}(V^{\ast})$, as $w^{\ast}_R(V^{\ast})$ is linear in $V$. Thus, the critical volume for the van der Waals equation is $V_{cr} = 3b$. One also has $w^{\ast}_L[1/(3b)] = 2a/(27b^2)$. The van der Waals equation must have a triple root at $V_{cr} = 3b$. Hence, at $V^\ast = 1/(3b)$, the straight line $w^{\ast}_R(V^{\ast})$ must have the same slope as the cubic $w^{\ast}_L(V^{\ast})$ and it must also pass through point $\left(1/(3b), 2a/(27b^2)\right)$. The slope of $w^{\ast}_L(V^{\ast})$ at $1/(3b)$ is its derivative at that point, namely, $a/(3b)$ and the equation of the tangent line through the inflection point of $w^{\ast}_L(V^{\ast})$ is $w^{\ast}_t(V^{\ast}) = aV^{\ast}/(3b) - a/(27 b^2)$. To determine the temperature and pressure the tangent line $w^{\ast}_t(V^{\ast})$ corresponds to, compare the respective coefficients of $w^{\ast}_t(V^{\ast}) = aV^{\ast}/(3b) - a/(27 b^2)$ and $w^{\ast}_R(V^{\ast}) = (R T + b p) V^\ast - p$. One immediately determines that $p_{cr} = a/(27b^2)$ and, hence, upon inserting this into $RT + bp = a/(3b)$, one gets $T_{cr} = 8a/(27Rb)$. \\
As $p$ decreases along an isotherm, the intersection point of $w^{\ast}_R(V^{\ast}) = (R T + b p) V^\ast - p$ with the ordinate moves in the direction of zero from below. As already seen, in the limit $V \to 2b^+$ (the double root) along the isotherm with temperature $T_0 = a/(4Rb)$, one has $p \to 0$. Then the straight line $w^{\ast}_R(V^{\ast})$ will tend to $R T_0 V^\ast$ --- the dashed line through the origin on Figure 2a, or the dashed curve through the origin, $w^{\ast}_L(V^{\ast}) - R T_0 V^\ast$, on Figure 2b. This curve is a limit curve. The curves that will be considered correspond to higher $T$. With the increase of $p$ and $T$, the straight line  $w^{\ast}_R(V^{\ast})$ exhibits both counter-clockwise rotation and downward translation (see Figure 2a). This straight line cannot intersect the cubic $w^{\ast}_L(V^{\ast})$ three times on one side of the inflection point $D$ which has abscissa $V^\ast = 1/(3b)$. One of the intersection points ($C$) is necessarily to the left of $D$. As the cubic $w^{\ast}_L(V^{\ast})$ is never negative for $V^\ast < 1/b$, intersection with the straight line $w^{\ast}_R(V^{\ast})$ is possible only when the graph of the straight line is above the abscissa. Hence, the localization interval of the smallest root $V^\ast_C \,$ is $p/(RT + bp) < V^\ast_C < 1/(3b)$. Thus, the biggest root of the van der Waals equation satisfies $3b \le V_C < RT/p + b = V_0 + b$, where $V_0 = RT/p$ is Clapeyron's ideal gas volume. \\
One of the other two intersection points of $w^{\ast}_L(V^{\ast})$ with $w^{\ast}_R(V^{\ast})$, point $A$, is necessarily to the right of the inflection point $D$, i.e. $V^\ast_A > 1/(3b)$ --- see Figure 2a. Point $A$ could be on either side of point $F$ --- the point at which the limiting straight line $R T_0 V^\ast$ is tangent to the cubic $w^{\ast}_L(V^{\ast})$. However, point $A$ cannot be to the right of point $H$ --- the point at the local maximum of the cubic $w^{\ast}_L(V^{\ast})$ --- that is, $V_A^\ast < 2/(3b)$, for the following reason. Consider an auxiliary straight line, parallel to the limiting straight line $R T_0 V^\ast$, and passing through the local maximum (point $H$) of $w^{\ast}_L(V^{\ast})$, that is through point $\left( 2/(3b), 4a/(27 b^2) \right)$. The ordinates of the points on this auxiliary straight line are given by  $[a/(4b)] V^\ast - a/(54b^2)$. This line is tangent to the cubic $w^{\ast}_L(V^{\ast})$ at point $V^{\ast} = 1/(6b)$. Any parallel straight line below this one will intersect $w^{\ast}_L(V^{\ast})$ only once --- to the right of the local maximum of $w^{\ast}_L(V^{\ast})$. In other words, if there are three intersection points, they necessarily exist for $V^\ast < 2/(3b)$. As $p$ increases from zero, the straight line $w^{\ast}_R(V^{\ast}) =  (R T + b p) V^\ast - p$ ``departs" from the dashed line $ - R T_0 V^\ast$ on Figure 2a --- not only by ``sliding" down, but also by rotating counter-clockwise. Hence, if the polynomial $w^{\ast}(V^{\ast})$ has three real roots, they are all to the left of point $H$ on Figures 2a and 2b. The localization interval of the biggest root $V^\ast_A$ is therefore $1/(3b) \le V^\ast_A < 2/(3b)$ and the smallest root of the van der Waals equation satisfies $3b/2 < V_A \le 3b$. \\
The middle root $V_B$ of the van der Waals equation could be on either side of the inflection point $D$ with abscissa $3b$ (see Figures 2a and 2b), but it is never smaller than $2b$ --- point $F$ on Figure 1. From above, $V_B$ is bounded by $V_C$, which, in turn and as already seen, is bounded from above by $RT/p + b = V_0 + b$. Thus the localization interval of $V_B$ is $2b < V_B < RT/p + b = V_0 + b$. \\
Excluding the unstable state $V_B$, the localization intervals for $V_A$ and $V_C$ are also their {\it isolation intervals} --- as one and only one root of the van der Waals equation is contained in each of them:  $3b/2 < V_A \le 3b$ and $3b \le V_C < RT/p + b = V_0 + b$.

\section{Isolation Intervals of the Local Extrema of $\,\boldsymbol{p(V)}$}

The localization interval of the middle root of the van der Waals equation can be narrowed down by the determination of the isolation  intervals of the local extrema of the curve $p(V)$, the knowledge of which is important in its own right in order to localize the unstable states of the van der Waals model. This can be done in a similar way and as follows. \\
The numerator of the derivative of $p(V)$ from (\ref{vdw2}) with respect to $V$ (at any constant $T$) is
\b
\label{pi}
\pi(V) = -R T V^3 + 2 a V^2 - 4 a b V + 2 a b^2.
\e
Using the same trick as before, one can determine the roots of equation $\pi(V) = 0$ as the reciprocals of the roots of
\b
\pi^{\ast}(V^{\ast}) = \pi^{\ast}_L(V^{\ast}) - \pi^{\ast}_R(V^{\ast}),
\e
where:
\b
\pi^{\ast}_L(V^{\ast}) & = &  2 a b^2 V^{\ast^3} - 4 a b V^{\ast^2}, \\
\pi^{\ast}_R(V^{\ast}) & = &  - 2 a V^\ast + RT.
\e
This time, it is the parameter $T$ that has been ``transferred" from the cubic to the straight line. The cubic $2 a b^2 V^{\ast^3} - 4 a b V^{\ast^2}$ is also ``fixed" by the parameters $a$ and $b$ only. By varying $T$, only the $y$-intercept of the negative-slope straight line $- 2 a V^\ast + RT$ varies. The graphs of $\pi^{\ast}_L(V^{\ast})$ and $\pi^{\ast}_R(V^{\ast})$ are shown on Figure 3a and the graph of $\pi^{\ast}(V^{\ast}) = \pi^{\ast}_L(V^{\ast}) - \pi^{\ast}_R(V^{\ast})$ is shown on Figure 3b. 

\vskip1cm
\begin{center}
\begin{tabular}{cc}
\includegraphics[width=67mm]{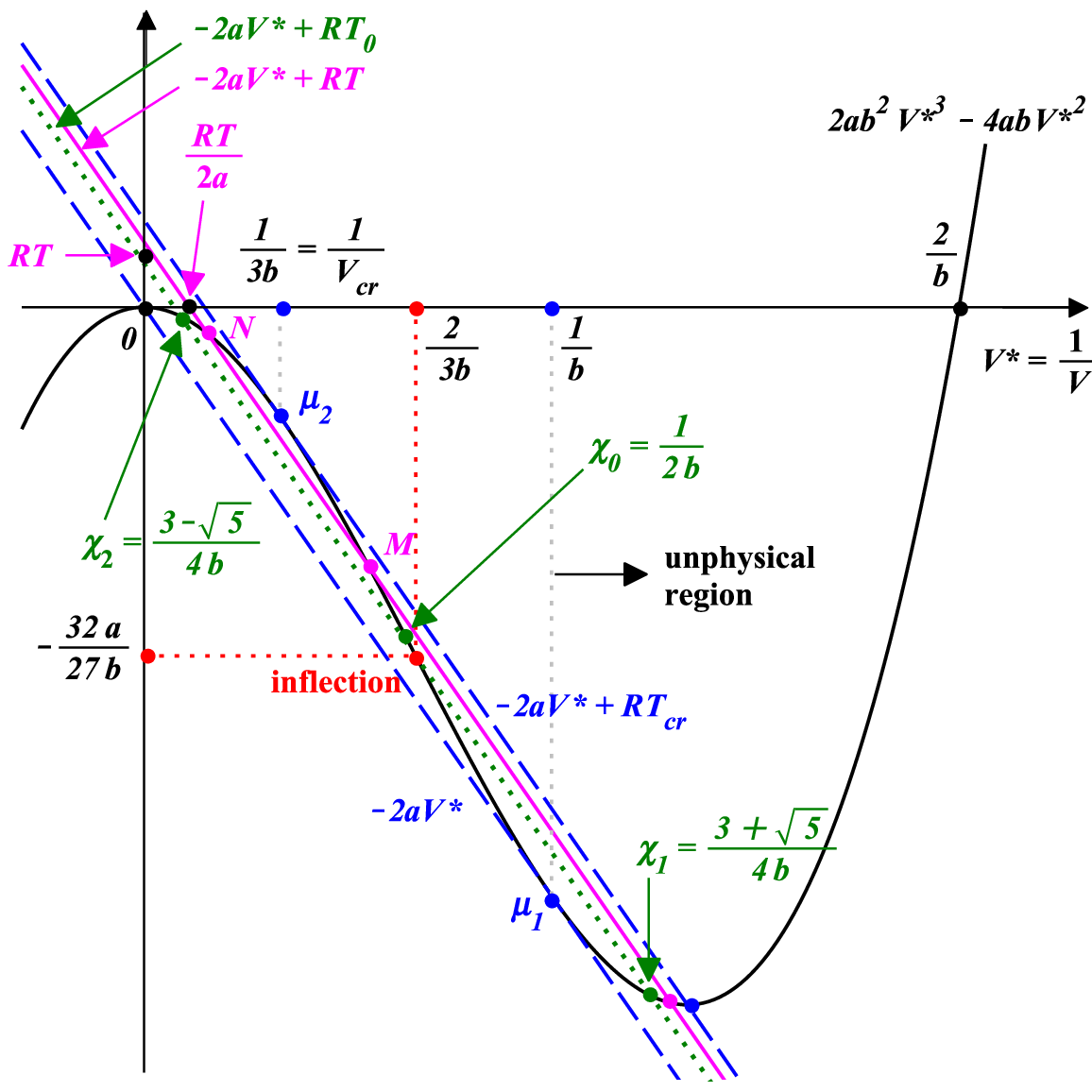} & \includegraphics[width=67mm]{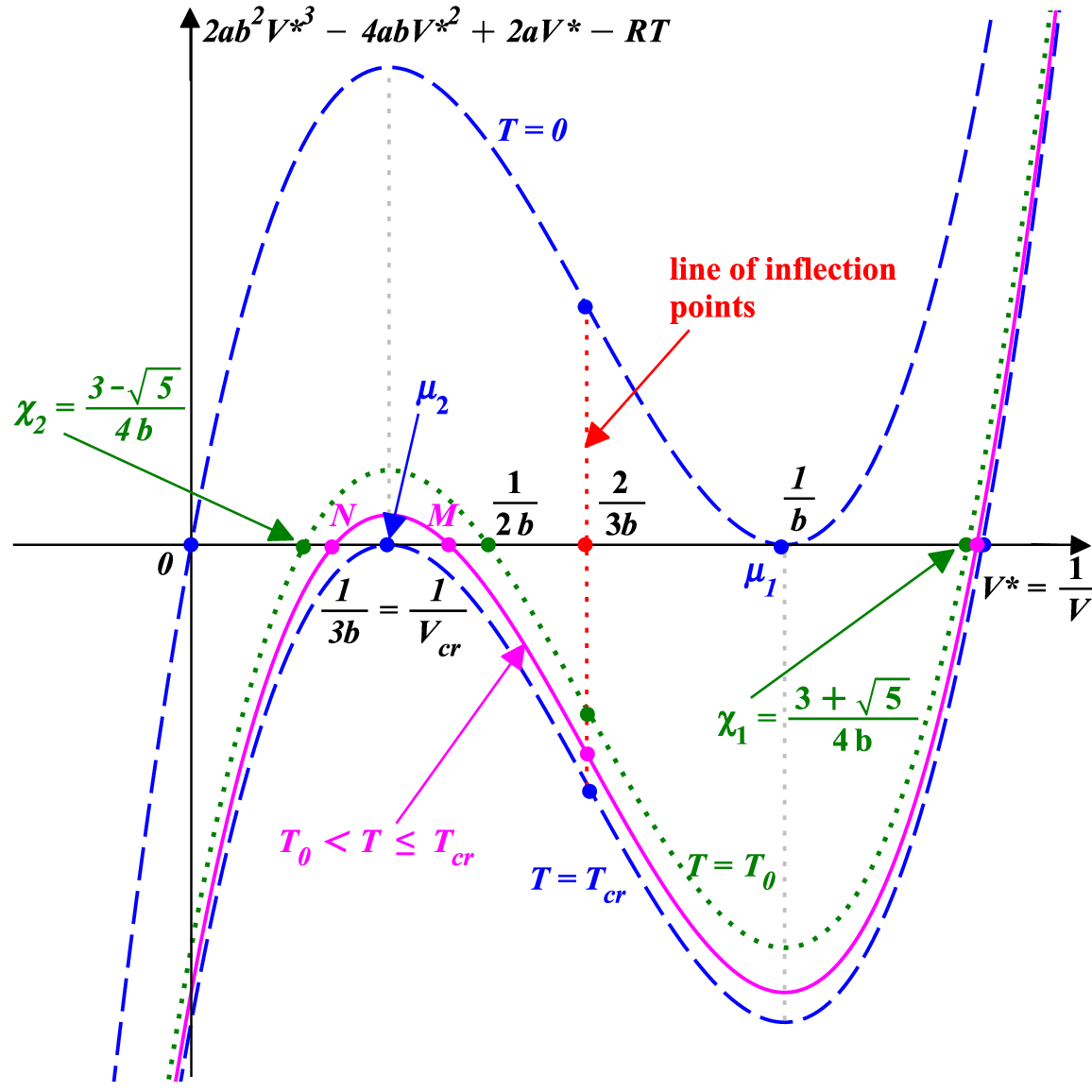} \\
{\scriptsize {\bf Figure 3a}} &  {\scriptsize {\bf Figure 3b}} \\
{\scriptsize {\bf The equation}} & {\scriptsize {\bf The polynomial}} \\
{\scriptsize {\bm{$a b^2 V^{\ast^3} - 4 a b V^{\ast^2} = - 2 a V^\ast + RT$}}} & {\scriptsize {\bm{$\pi^\ast(V^\ast) = a b^2 V^{\ast^3} - 4 a b V^{\ast^2} + 2 a V^\ast - RT$}}} \\
& \\
\end{tabular}
\end{center}

\noindent
One can pose the question: at what values of $V^\ast$ are the graphs of $\pi^{\ast}_L(V^{\ast})$ and $\pi^{\ast}_R(V^{\ast})$ tangent to each other, that is, when does  the full $\pi^{\ast}(V^{\ast})$ have a double root $\mu$? One can ``slide" the full $\pi^{\ast}(V^{\ast})$ up and down by varying its free term $RT$ (Figure 3a) and, at two certain values of $T$, this will be realized. In terms of the two ``parts" $\pi^{\ast}_L(V^{\ast})$ and $\pi^{\ast}_R(V^{\ast})$ on Figure 3a, only the latter is sensitive to the temperature. When the full $\pi^{\ast}(V^{\ast})$ has a double root $\mu$, $\pi^{\ast}_R(V^{\ast})$ is tangent to $\pi^{\ast}_L(V^{\ast})$ at point $\mu$. This implies equality of the two functions and equality of their derivatives at that point. Hence, from the latter, $\mu$ are the roots of $(d/dV)\pi^{\ast}_L(V^{\ast}) = (d/dV) \pi^{\ast}_R(V^{\ast})$. This is the quadratic equation $3 ab^2 V^{\ast^2} - 4 ab V^\ast + a = 0$ and its roots are $\mu_1 = 1/b$ and $\mu_2 = 1/(3b)$. From the former, one also has $\pi^{\ast}_L(\mu_{1,2}) = \pi^{\ast}_R(\mu_{1,2})$ and therefore $R T_{1,2} = 2 a b^2 \mu_{1,2}^3 - 4 a b \mu_{1,2}^2 + 2 a \mu_{1,2}$. One immediately gets that $T_2 = T_{cr} = 8a/(27Rb^2)$ and $T_1 = 0$. Hence, two limiting straight lines appear: $- 2 a V^\ast + RT_{cr}$ and $- 2 a V^\ast$. These are represented by the dashed lines on Figure 3a. The corresponding ``full" limiting curves on Figure 3b are the dashed $2 a b^2 V^{\ast^3} - 4 a b V^{\ast^2} +  2 a V^\ast - RT_{cr}$ and $2 a b^2 V^{\ast^3} - 4 a b V^{\ast^2} +  2 a V^\ast$, which passes through the origin. Noting next that $T$ is considered greater than $T_0$, the realizable straight lines on Figure 3a are those between the dotted line $- 2 a V^\ast + RT_{0}$ and the upper dashed line $- 2 a V^\ast + RT_{cr}$. On Figure 3b, the realizable curves are those between the lower dashed curve $2 a b^2 V^{\ast^3} - 4 a b V^{\ast^2} +  2 a V^\ast - RT_{cr}$ and the dotted curve $2 a b^2 V^{\ast^3} - 4 a b V^{\ast^2} +  2 a V^\ast - RT_{0}$. Note next that the inflection point of $\pi^{\ast}_L(V^{\ast})$ is at $V^\ast = 2/(3b)$. The dotted line (corresponding to $T = T_0$) intersects  $\pi^{\ast}_L(V^\ast)$ at $V^\ast = 1/(2b)$ --- to the left of inflection point $V^\ast = 2/(3b)$ --- see Figure 3a.  \\
When $T_0 < T < T_{cr}$, the discriminant $-4 a^2 R T b^3 (27 R T b - 8a)$ of $\pi(V)$ is positive. Hence, $p(V)$ has three extrema. One of them is in the unphysical region $0 < V < b$, while the isolation  intervals of the other two extrema --- the local minimum at point $M$ and the local maximum at point $N$ --- can be immediately read from Figures 3a and 3b as follows. The local minimum of $p(V)$ is at point $M$ whose abscissa is between $\mu_2 = 1/(3b)$ and the middle intersection point $\chi_0 = 1/(2b)$ of $2 a b^2 V^{\ast^3} - 4 a b V^{\ast^2}$ with the dotted line $- 2 a V^\ast + RT_0$. The local maximum of $p(V)$ is at point $N$ whose abscissa is between the left-most intersection point $\chi_2 = (3 - \sqrt{5})/(4b)$ of $2 a b^2 V^{\ast^3} - 4 a b V^{\ast^2}$ with the dotted line $- 2 a V^\ast + RT_0$ and $\mu_2 = 1/(3b)$. [The right-most intersection point of $2 a b^2 V^{\ast^3} - 4 a b V^{\ast^2}$ with the dotted line $- 2 a V^\ast + RT_0$ is at $\chi_1 = (3 + \sqrt{5})/(4b)$ --- in the unphysical region $0 < V < b$.] \\
Therefore, the isolation  intervals of the two extrema of $p(V)$ (which exist for $T_0 < T < T_{cr}$) are as follows. The local minimum $M$ is at volume $V_M$ such that $2b < V_M < 3b$. The local maximum $N$ is at volume $V_N$ such that $3b < V_N < (3 + \sqrt{5})b  \approx 5.236b$.  \\
The unstable states of the van der Waals model are therefore between the lower bound of the local minimum and the upper bound of the local maximum, that is, they lie in an interval within the interval $\left(2b, (3 + \sqrt{5})b \right)$. The root $V_B$ of the van der Waals equation now has a narrower localization interval: $2b < V_B < (3 + \sqrt{5})b $. \\
Consider now a van der Waals equation for which one is allowed to take temperatures below $T_0 = 1/(4Rb)$. Clearly, the pressure $p$ on the $p\!-\!V$ diagram will exhibit a ``dip" below the abscissa. Then, the limiting dotted line $- 2 a V^{\ast} + R T_0$ on Figure 3a must be replaced by a new limiting straight line --- one that is below the dotted one and above the straight line $- 2 a V^{\ast}$, which corresponds to $T = 0$. The equation of this new limiting straight line will be $- 2 a V^{\ast} + R \, \tau$, where $\tau$ --- the new minimum temperature of the model --- is between 0 and $T_0 = 1/(4Rb)$. This straight line crosses the abscissa at $V^{\ast} = R \, \tau/(2a)$. Hence, the intersection points of $- 2 a V^{\ast} + R \tau$ with the cubic $2 a b^2 V^{\ast^3} - 4 a b V^{\ast^2}$ can only happen to the right of point $V^{\ast} = R \, \tau/(2a)$. The first intersection is between $R \, \tau/(2a)$ and $\mu_2 = 1/(3b)$, the second --- between $\mu_2 = 1/(3b)$ and $\mu_1 = 1/b$, and the third intersection is in the unphysical region of $V^{\ast} > 1/b$ --- see Figure 3a. With this, one immediately sees that the local minimum of $p(V)$ now has isolation interval $(b, 3b)$, while the local maximum has isolation interval $(3b, 2a/(R \, \tau))$. The root $V_B$ of the van der Waals equation now has the localization interval $b < V_B < 2a/(R \, \tau)$ and the unstable states of the van der Waals equation fall into this interval too. As, for $\tau \le T < T_{cr}$, the volume $V_A$ is smaller than $V_B$, then the new localization interval of $V_A$ will be $(b, 3b)$.  \\
When $T > T_{cr} = 8a/(27Rb)$, the discriminant $-4 a^2 R T b^3 (27 R T b - 8a)$ of $\pi(V)$ is negative and $p'(V)$ has only one real root. This root is beyond $\mu_1$ --- in the unphysical region $0 < V < b$. Such isotherms do not exhibit minima and maxima that are physically accessible and hence they can be intersected by an isobar only once --- at a unique positive value of $V$ for each positive value of $p$ (point $E$ on Figures 1, 2a, and 2b). For temperatures $T > T_{cr}$ and for pressures $p \ge p_{cr}$, the volume $V_E$ at point $E$ is bound by $b$ from below --- happening, clearly, at any temperature when $p$ is infinitely large (see Figure 1). The volume $V_E$ is bound from above by $b + RT/p$ --- the intersection of $(RT + bp) V^\ast - p$ and the abscissa $V^\ast$ --- see Figure 2a. The maximum of the upper bound $b + RT/p$, for any finite $T > T_{cr}$ and for $p \ge p_{cr}$, is $b + RT/$min$(p)= b + RT/p_{cr} = b + (27Rb^2/a)T$, thus $b < V_E <  b + (27Rb^2/a)T$ in this case.  For temperatures $T > T_{cr}$ and for pressures $p < p_{cr}$, the unique intersection point $E$ of the isobar and the isotherm will be bound from below by $b + (27Rb^2/a)T$ and above by $b + RT/p = b + V_0$ (see Figure 1).

\section{On Finding the Roots of the van der Waals Equation}
Consider next how the three real roots $V_A, V_B$, and $V_C$ of the cubic van der Waals equation can be found for any chosen temperature $\widetilde{T}$ between $T_0 = a/(4Rb)$ and $T_{cr} = 8a/(27Rb)$. \\
The equality of the two areas bounded by the isotherm $T = \widetilde{T}$ and the isobar $p = \widetilde{p}$ means that $\int_{V_A}^{V_B} \, [ \widetilde{p} - p(V)] \, dV = \int_{V_B}^{V_C} \, [ p(V) - \widetilde{p}] \, dV$. By substituting (\ref{vdw2}), one gets \cite{b}:
\b
\label{int}
\int\limits_{V_A}^{V_B} \left( \widetilde{p} - \frac{R \widetilde{T}}{V-b} + \frac{a}{V^2} \right) dV = \int\limits_{V_B}^{V_C} \left(\frac{R \widetilde{T}}{V-b} - \frac{a}{V^2} - \widetilde{p} \right) dV.
\e
Performing the integration and solving for $\widetilde{p}$ yields \cite{b}
\b
\label{V1}
\widetilde{p} = \frac{R \widetilde{T}}{V_C - V_A} \, \ln \frac{V_C - b}{V_A - b} - \frac{a}{V_A V_C}.
\e
The three real roots $V_A, V_B$, and $V_C$ of the van der Waals equation
\b
p V^3 - (R T + b p) V^2 + a V - ab = 0
\e
are related through the Vi\`ete formul\ae:
\b
V_A + V_B + V_C & = & b + \frac{RT}{p}, \\
V_A V_B + V_A V_C + V_B V_C & = & \frac{a}{p}, \\
V_A V_B V_C & = & \frac{ab}{p}.
\e
To determine the three unknowns $\widetilde{p}$, $V_A$, and $V_C$ in (\ref{V1}), while treating $\widetilde{T}$ as given, Boltzmann suggests \cite{b} to use (\ref{V1}), together with the following two conditions: that $V_A$ is the smallest root of the van der Waals equation and that $V_C$ is the largest one. This can be done as follows. Expressing $V_B + V_C$ as $b + R \widetilde{T} / p - V_A$, using the first Vi\`ete formula, and $V_B V_C$ as $ab/(p V_A)$, using the third one, the van der Waals equation can be re-written as
\b
\label{V2}
p \, V^3 - (R \widetilde{T} + b p) V^2 + \,\, & & \!\!\!\!\!\!\!\!\!\!\!\!\! a V - ab  \,\,\, =  \,\,\, p \, (V - V_A)(V - V_B)(V-V_C) \nonumber \\
& \!\!\!\!\! = & p \, (V - V_A)\left[V^2 - \left( b + \frac{R \widetilde{T}}{p} - V_A \right) V + \frac{ab}{p V_A} \right] = 0.
\e
The roots $V_{B}$ and $V_{C}$ of the van der Waals equation are the zeros of the quadratic polynomial $V^2 - (b + R \widetilde{T}/ p - V_A)V + ab/(p \, V_A)$ in the square brackets of (\ref{V2}). This polynomial vanishes when $V = V_B$ and when $V = V_C$ and in either of these cases, one has $p = \widetilde{p}$. Thus, these two roots are:
\b
\label{vcb}
V_{C, B} = \frac{-(V_A - b) \widetilde{p} + R \widetilde{T} \pm \sqrt{(V_A -b)^2 \widetilde{p}^{\, 2} + (-2R \widetilde{T} V_A + 2Rb \widetilde{T} - \frac{4ab}{V_A})\widetilde{p} + R^2 \widetilde{T}^2}}{2\widetilde{p}}.
\e
Observe that, from (\ref{vdw2}), one has $\widetilde{p} = (R \widetilde{T} V_A^2 - a V_A + ab)/[V_A^2(V_A - b)]$ when $V = V_A$. Hence substitute this $\widetilde{p}$ into the left-hand side of (\ref{V1}) and also into (\ref{vcb}). By doing so, $\widetilde{p}$ is eliminated from (\ref{V1}) and (\ref{vcb}). Then substitute the obtained in this way from (\ref{vcb}) $V_C$ into (\ref{V1}). This yields an equation for the smallest root of the van der Waals equation, the stable $V_A$, for the selected temperature $\widetilde{T}$. Once $V_A$ is determined from (\ref{V1}), one immediately finds $\widetilde{p} = (R \widetilde{T} V_A^2 - a V_A + ab)/[V_A^2(V_A - b)]$ explicitly for the chosen value $\widetilde{T}$. Finally, the remaining two positive and greater roots $V_B$ and $V_C$ of the van der Waals equation are then easily obtained from (\ref{vcb}) when one substitutes into it the already determined $V_A$ and $\widetilde{p}\, $ for the chosen $\widetilde{T}$. \\
One should note that the above equations are transcendental and finding the solutions algebraically is impossible --- only approximate solutions can be found.

\section{Discussion and Conclusions}
The results reported allow one to trace the isothermal compression of {\it any substance} subjected to the van der Waals model at any temperature above $T_0$ (only positive pressures are considered), starting at the gaseous phase at high volumes, in the following manner. The localization interval of the biggest van der Waals root $V_C$ is $3b \le V_C < V_0 + b$, where $V_0 = RT/p$ is the ideal gas volume. Hence, when the volume reaches a value somewhere below $V_0 + b$, the phase of supercooled vapour commences (point $C$ on Figure 1). This phase continues until the local maximum  (point $N$ on Figure 1) is reached, which cannot happen for volumes above $(3 + \sqrt{5})b$ --- the upper end of the isolation  interval of the local maximum $N$. The states between points $N$ and $M$ (the local maximum and the local minimum, respectively, see Figure 1) are unstable and cannot be realized as the pressure drops with the drop of the volume. This is the phase in which both gas and liquid co-exist. With the further compression of the substance beyond the local minimum $M$, only superheated liquid exists. As the lower end of the isolation  interval of the local minimum $M$ is determined as $2b$, one can be certain that there is no gas at volume $V < 2b$ --- there is only superheated liquid. Once the volume gets below $3b/2$ --- the lower end of the localization interval of the smallest root $V_A$ --- the phase is entirely that of liquid (compression below $b$ is impossible in the van der Waals model).


\end{document}